\documentstyle{mn}

\newcounter{parentequation}
\def\beglet{
  \addtocounter{equation}{1}%
  \setcounter{parentequation}{\value{equation}}%
  \setcounter{equation}{0}%
  \def\theequation{\arabic{parentequation}\alph{equation}}%
  \ignorespaces
}
\def\endlet{
  \setcounter{equation}{\value{parentequation}}%
  \def\theequation{\arabic{equation}}%
}

\def\ltsima{$\; \buildrel < \over \sim \;$}
\def\gtsima{$\; \buildrel > \over \sim \;$}
\def\simlt{\lower.5ex\hbox{\ltsima}}
\def\simgt{\lower.5ex\hbox{\gtsima}}

\def\etal{{\it et al.}\rm}
\def\etals{{\it et al. }\rm}

\begin{document}

\title[CMB Quadrupole and Spatial Curvature]
{Is the Low CMB Quadrupole a Signature of Spatial
Curvature?}

\author[G. Efstathiou]{G. Efstathiou\\
Institute of Astronomy, Madingley Road, Cambridge, CB3 OHA.}

\maketitle

\begin{abstract}
The temperature anisotropy power spectrum measured with the Wilkinson
Microwave Anisotropy Probe (WMAP) at high multipoles is in spectacular
agreement with an inflationary $\Lambda$-dominated cold dark matter
cosmology. However, the low order multipoles (especially the
quadrupole) have lower amplitudes than expected from this cosmology,
indicating a need for new physics. Here we speculate that the low
quadrupole amplitude is associated with spatial curvature. We show that
positively curved models are  consistent with the WMAP data
and that the quadrupole amplitude can be reproduced if the 
primordial spectrum truncates on scales comparable to 
the curvature scale.

\vskip 0.1 truein

\noindent
{\bf Key words}: cosmic microwave background, cosmology.

\vskip 0.1 truein

\end{abstract}

\section{Introduction}

The recent measurements of the cosmic microwave background (CMB)
anisotropies by  WMAP
(Bennett \etals 2003; Spergel \etals 2003; Peiris \etals 2003) are in
striking agreement with a `concordance' cosmology (hereafter referred
to as $\Lambda$CDM) that has been built up over the last few years
using many different data sets (see {\it e.g.}  Bahcall \etals 1999,
for a review). According to this model, the Universe is spatially flat
with adiabatic, nearly scale invariant initial fluctuations as
predicted in simple inflationary models (see {\it e.g.} Liddle and
Lyth 2000).  In addition, the present day Universe is dominated by a cosmological
constant, consistent with the magnitude-redshift relation of Type Ia
supernovae (Reiss \etals 1998; Perlmutter \etals 1999) and the
large-scale distribution of galaxies (Efstathiou \etals 1990, 2002).

The high precision of the WMAP experiment leads to tight constraints
on various cosmological parameters. For example, from the WMAP data alone
Spergel \etals (2003) find that the best fit spatially flat $\Lambda$CDM
models have a Hubble parameter of $h=0.72 \pm 0.05$
\footnote{Here $h$ is Hubble's constant $H_0$ in units of $100{\rm
km}{\rm s}^{-1} {\rm Mpc}^{-1}$. The cosmic densities in baryons, cold
dark matter and vacuum energy are denoted by $\Omega_b$, $\Omega_c$
and $\Omega_\Lambda$. The total matter density is $\Omega_m = \Omega_b
+ \Omega_c$ and the curvature is fixed by $\Omega_k = 1 - \Omega_m -
\Omega_\Lambda$.},  scalar
spectral index of $n_s = 0.99 \pm 0.04$ and  physical baryonic and cold dark
matter densities of $\omega_b = \Omega_bh^2 = 0.024 \pm 0.001$ and
$\omega_c = \Omega_c h^2 = 0.12 \pm 0.02$  (all $1\sigma$
errors). Furthermore, from the temperature-polarization cross
power-spectrum the WMAP data suggest a high optical depth for
secondary reionization, $\tau =  0.17 \pm 0.04$, indicating significant
star formation at high redshifts ({\it e.g.} Wyithe and Loeb 2003;
Cen 2003;  Haiman and Holder 2003).

There is, however, a potential problem with the simple concordance
model. The WMAP results confirm the low amplitude of the CMB
quadrupole seen by COBE (Hinshaw \etals 1996). As Bennett \etals
(2003) comment, the amplitude of the quadrupole (and to a lesser
extent the octopole) is low compared with the predictions of
$\Lambda$CDM models that otherwise fit the rest of the power spectrum
to extraordinarily high precision. The WMAP team present a convincing
case that the low CMB multipoles are not significantly affected by
foreground Galactic emission (see also Tegmark, de Oliveira-Costa and
Hamilton 2003).  The discrepancy is evident in a particularly dramatic
form in the temperature angular correlation function, which shows an
almost complete lack of signal on angular scales $\simgt 60$ degrees.
According to Spergel \etals (2003), the probability of finding such a
result in a spatially flat $\Lambda$CDM cosmology is about $1.5 \times
10^{-3}$. This is small enough to be worrying and may indicate the
need for new physics. Since this paper was submitted, a number of
authors have questioned the interpretation of Spergel {\it
et al.}'s result (Gazta\~naga \etals 2003; Tegmark \etals 2003, Bridle
\etals 2003, Efstathiou 2003). The analysis of Efstathiou (2003) suggests
that a more realistic probability of finding the observed quadrupole 
and octopole amplitudes in the concordance $\Lambda$CDM cosmology is
more like $0.05$. The data are suggestive of new physics, but 
not at a high level of statistical significance. Nevertheless, it is 
worth investigating models incorporating new physics and to identify any 
distinctive predictions (apart from low quadrupole and octopole amplitudes)
that they might make.

What sort of new physics might be required? One possibility is to
invoke fluctuations in a quintessence-like scalar field which can
introduce features on scales comparable to the present day Hubble
radius (Caldwell, Dave and Steinhardt 1998). The difficulty here is to
cancel the large integrated Sachs-Wolfe anisotropy associated with a
high vacuum energy density (see {\it e.g.} DeDeo, Caldwell and
Steinhardt 2003). Here we speculate that the low quadrupole is linked
to the curvature scale. As discussed in Section 2, although the WMAP
data are consistent with a spatially flat Universe, models with
significant positive curvature are allowed (Spergel \etals
2003). Since we have no accepted theory to explain a closed universe
(see Linde 1995; Gratton, Lewis and Turok, 2002; Ellis \etals 2002;
Uzan, Kirchner and Ellis 2003; Linde 2003; Contaldi \etals 2003 for
some ideas) it follows that we do not have a firm theory for the shape
of the fluctuation spectrum. We therefore propose in Section 3 that
the low quadrupole and octopole amplitude measured by WMAP may be
indicative of a truncation in the primordial power spectrum on the
curvature scale.\footnote{A closed inflationary model that can produce
such a truncation is discussed by Contaldi \etals 2003.}

\section{Constraints on Spatial Curvature and the Geometrical Degeneracy}

It is well known that parameters estimated from the CMB anisotropies
show strong degeneracies ({\it e.g.}  Bond \etals 1997; Zaldarriaga,
Spergel and Seljak 1997; Efstathiou and Bond 1999). The best known is
the geometrical degeneracy between the matter density, vacuum energy
and curvature. Models will have nearly identical CMB power spectra if
they have identical initial fluctuation spectra and reionization
optical depth, and if they have identical values of $\omega_b$,
$\omega_c$ and acoustic peak location parameter \beglet
\begin{eqnarray}
&& {\cal R} = {\Omega_m^{1/2} \over \vert \Omega_k\vert ^{1/2}}\; {\rm sin}_K
\left [ \vert \Omega_k\vert ^{1/2}   y \right ],  \label{eq:1a} 
\end{eqnarray}
where
\begin{eqnarray}
&& y  = \int_{a_r}^{1} {da \over [\Omega_m a + \Omega_ka^2 + 
\Omega_\Lambda a^4 ]^{1/2}}, \label{eq:1b}
\end{eqnarray}
\begin{eqnarray}
&& \sin_K  = \left\{ \begin{array}{ll}
       {\rm sinh}  & \quad \Omega_k > 0 \\
       {\rm sin} & \quad \Omega_k < 0
      \end{array}
\right .  , \label{eq:1c}
\end{eqnarray}
\endlet
and $a_r$ is the scale factor at recombination normalised to unity at the present
day. 

The geometrical degeneracy is almost exact and precludes reliable
estimates of either $\Omega_\Lambda$ or the Hubble parameter $h$ from
measurements of the CMB anisotropies alone. As an example, Figure 1
shows the temperature power spectrum measured by WMAP together with a
scale-invariant $\Lambda$ CDM model with $\omega_b$, $\omega_c$, $h$
and $\tau$ fixed to the WMAP best fit values quoted in the
Introduction.  The other curves in the figure show nearly degenerate
models with $\Omega_k = -0.05$, $-0.10$ and $-0.20$ with parameters
listed in Table 1. All of these models have been computed using the 
CMBFAST code of Seljak and Zaldarriaga (1996).

\begin{table}
\bigskip

\centerline{\bf \ \ \  Table 1:  
Parameters for degenerate models}

\begin{center}

\begin{tabular}{ccccc} \hline \hline
\smallskip 
 $\Omega_k$ & $\Omega_b$& $\Omega_c$ & $\Omega_\Lambda$ & h \cr
 & & & &    \cr
$\;\;0.00$ & $0.0463$ & $0.2237$ & $0.73$ & $0.720$ \cr
$-0.05$ & $0.0806$ & $0.3894$ & $0.58$ & $0.546$ \cr
$-0.10$ & $0.1114$ & $0.5386$ & $0.45$ & $0.446$ \cr
$-0.20$ & $0.1714$ & $0.8286$ & $0.20$ & $0.374$ \cr
\hline
\end{tabular}
\end{center}
\noindent
Note: These models have identical values of $\omega_b$ and $\omega_c$
fixed to the best fit values from WMAP.
\end{table}

The theoretical power spectra plotted in Figure 1 are, by
construction, identical at high multipoles and differ by less than the
cosmic variance (indicated by the error bars in Figure 1) at low
multipoles. They are therefore statistically indistinguishable using
CMB data alone. Closed models with large values of $\vert \Omega_k
\vert$ require low values of the Hubble constant and conflict with the
direct measurement reported by the HST key project team (Freedman
\etals 2001). Nevertheless, closed models with $\Omega_k \sim -0.05$
can be adjusted to have values of the cosmological parameters
consistent with other data. In fact, Spergel \etals (2003) find that
the best fit model to WMAP combined with other data sets is slightly
closed with $\Omega_k = -0.02 \pm 0.02$. However, as Figure 1
demonstrates, this value and the error depend critically on the
accuracy and interpretation of more complex `astrophysical' ({\it
i.e.} non-CMB) data. We conclude that closed models are consistent
with, and may even be marginally favoured by, the WMAP data.

\begin{figure}

\vskip 2.8 truein

\includegraphics{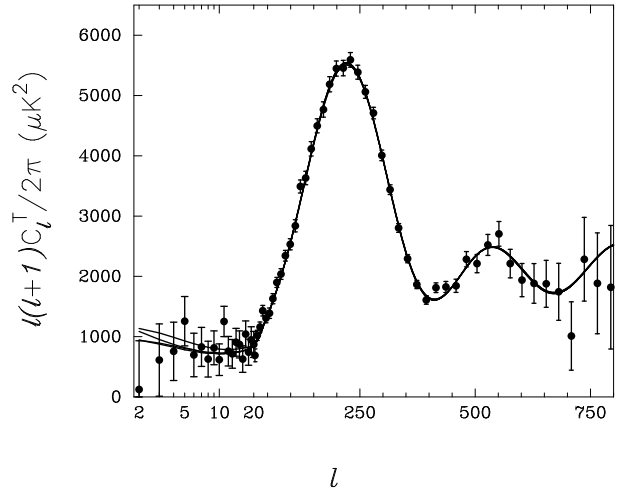}

\caption
{Illustration of the geometrical degeneracy.The filled circles show the
WMAP temperature power spectrum estimates. The nearly degenerate
models with $\Omega_k=0.0$, $-0.05$, $-0.10$ and $-0.20$ (see Table 1)
are shown by the solid lines and are identical except for minor
differences at low multipoles.  The error bars on the WMAP points use
the theoretical power spectrum for the $\Omega_k=0$ model.}

\label{figure1}

\end{figure}

\section{Low CMB multipoles and Spatial Curvature}

The metric of a Friedmann-Robertson-Walker model can be written in terms
of a development angle $\chi$ as
\begin{equation}
 ds^2 = dt^2 - a^2(t) \vert K\vert ^{-1} \left [  d\chi^2 + \sin^2_K\chi \left(
d\theta^2 + \sin^2 \theta d\phi^2 \right ) \right ],
\end{equation}
where $K = - H_0^2 \Omega_k/c^2$ defines a curvature scale $R_c =
\vert K \vert^{-1/2} = (c/H_0) \vert \Omega_k\vert^{-1/2}$. Linear perturbation theory of FRW
models with arbitrary curvature is well developed (see for example,
Wilson 1983; Abbott and Schaefer 1986; Lyth and Woszczyna 1995; White
and Scott, 1996; Zaldarriaga, Seljak and Bertschinger 1998; Zaldarriaga
and Seljak 2000).
As is well known, in a closed FRW model the eigenvalues $\beta$ of the
the Laplacian are discrete and related to physical comoving wavenumber $k$
by
\begin{equation}
 \beta^2 = ( 1 + k^2R_c^2).
\end{equation}
Modes with $\beta=1$ and $2$ are pure gauge modes (Abbott and Schaefer
1986) and so the complete spectrum of modes extends from $\beta = 3$ to infinity.

The mode spectrum in a closed universe therefore contains a characteristic
scale -- the curvature scale $R_c$. In the absence of a detailed model
for the origin of fluctuations in a closed universe it is not obvious how to
generalise the concept of `scale-invariant' fluctuations on scales
comparable to the curvature scale. For example,  if the potential fluctuations
are assumed to be constant per logarithmic interval in wavenumber $k$,  the
initial density fluctuation spectrum in the CMBFAST code must be set to
\beglet
\begin{equation}
 P_\rho (\beta) \propto   {(\beta^2 - 4)^2 \over \beta (\beta^2 - 1)},
\end{equation}
(see {\it e.g.} White and Bunn 1995). This form of the power spectrum
was used to compute the theoretical models plotted in Figure 1.
Starobinsky (1996) argues for equation (4a) based on slow roll
inflation and adopting the conformal vacuum as an initial
state. However, the initial conditions of this model are not understood in
terms of a fundamental theory.

In the absence of a well-motivated theory, we take the
CMB data at face value and investigate a power spectrum that truncates on
scales comparable to the curvature scale. For example, the heuristic form
\begin{equation}
 P_\rho (\beta) \propto  =  {(\beta^2 - 4)^2 \over \beta (\beta^2 - 1)}
\left [ 1 - {\rm exp} \left (-{ (\beta-3) \over 4} \right ) \right ],
\end{equation}
\endlet
damps the spectrum at low values of $\beta$.
The consequences for the CMB anisotropies are shown in Figure 2. The
WMAP temperature power spectrum at low multipoles is shown in the
upper panel and the TE temperature-E mode polarization spectrum is shown in
the lower panel.  The closed universe models from Table 1 with the
heuristic form of the power spectrum of equation (4b) are also
plotted. The important point to note from this Figure is that a 
truncation of the primordial power spectrum at $\beta \simlt 5$
can produce a significant change to the low order multipoles
even for values of $\Omega_k$ as small as $-0.05$, which are
difficult to rule out observationally. The low amplitudes
of the CMB quadrupole and octopole may therefore be linked
to the curvature scale.

Spergel \etals (2003), Tegmark \etals (2003) and Uzan \etals (2003)
comment that the low CMB multipoles may be an indication of a discrete
spectrum in a universe of finite size and non-trivial topology. The models
of this Section show that a periodic universe is not necessary if
the initial fluctuation spectrum truncates at around the curvature
scale.

\begin{figure}

\vskip 4.8 truein

\includegraphics{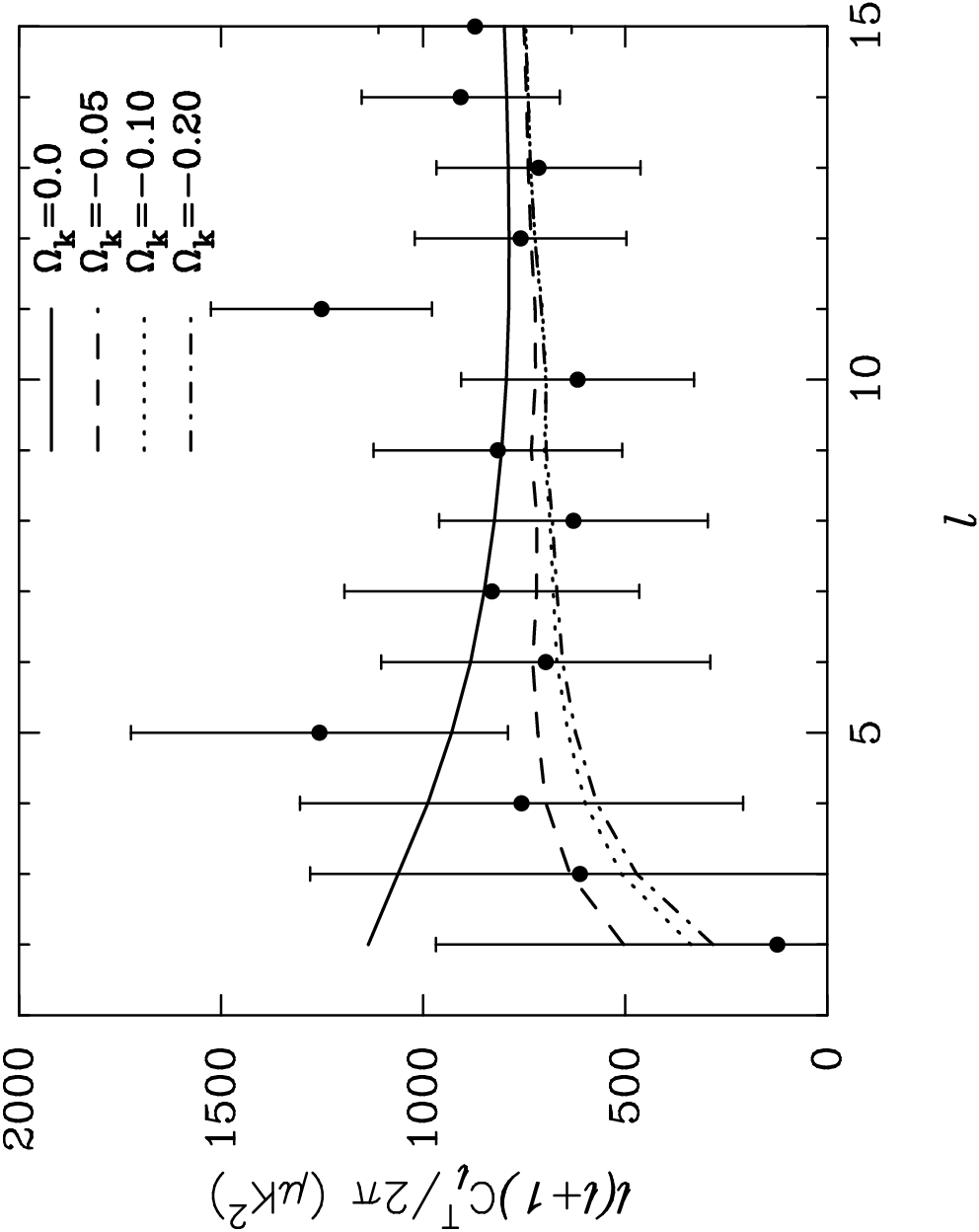}

\includegraphics{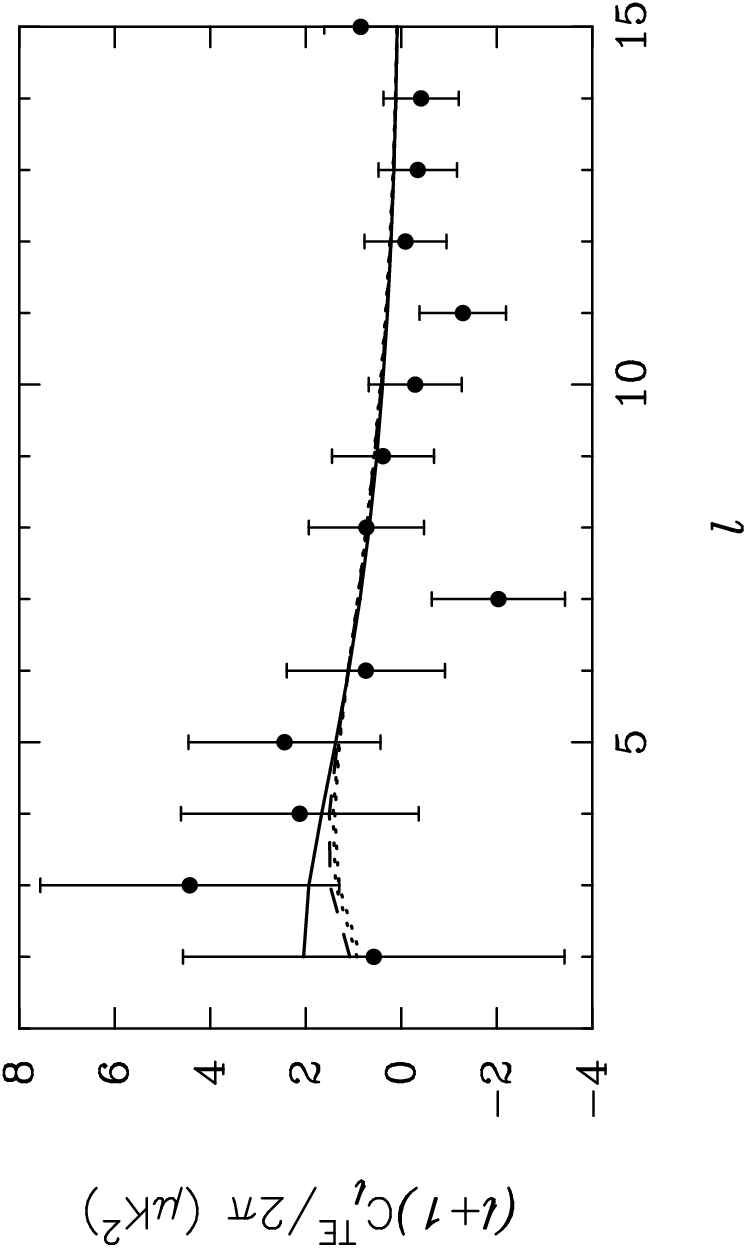}

\caption
{Low order CMB multipoles measured by WMAP: upper panel shows the temperature 
power spectrum and lower panel shows the temperature-polarization cross
power-spectrum. The solid line shows the reference $\Omega_k=0.0$ model 
which has been used to calculate the error bars. The dashed, dotted and
dot-dashed lines show the results for the positively curved models
with the heuristic spectrum of equation (4b).}

\label{figure2}
\end{figure}

\section{Conclusions}

The low amplitude of the quadrupole  may be the first tentative indication
of discordance with the otherwise remarkably successful $\Lambda$CDM
cosmology. It is worth taking seriously as a possible pointer towards
new physics. 

Although CMB data are consistent with a spatially flat universe, the
geometrical degeneracy makes it impossible to constrain $\Omega_k$
accurately from CMB data alone. To pin down $\Omega_k$ to an accuracy
of $0.05$ or better requires more messy `astrophysical' data, for
example, galaxy clustering or direct measurements of the Hubble
constant. Unless one has a good understanding of the systematic errors
in these more complicated observations, it will not be easy to rule
out models with a small positive curvature. Nevertheless, it may be 
possible with future experiments to constrain $\Omega_k$ to this
kind of accuracy.

A closed Universe possesses a charactertic curvature scale $R_c =
(c/H_0) \vert \Omega_k\vert^{-1/2}$. The proposal explored in this
note is that the low CMB quadrupole amplitude may be related to a
truncation of the primordial fluctuation spectrum on the curvature
scale. If this, admittedly speculative, proposal were right, we would
need to add the small positive curvature to the list of `cosmic
coincidences' (most notably the small value of $\Lambda$) that plague
modern cosmology. Furthermore, a positive cosmological curvature would
require a radical shift from the simple inflationary paradigm, which
many cosmologists might feel is too high a price to pay to explain two
or three of the thousand or more CMB multipoles. However,
discrepancies between theory and observations must be taken seriously,
wherever and whenever they occur.

\vskip 0.1 truein

\noindent
{\bf Acknowledgments:} I thank Sarah Bridle, Anthony Challinor, Antony Lewis,
Martin Rees and Jochen Weller  for helpful comments.

\end{document}